\documentclass{appolb}
\usepackage{graphicx}
\usepackage{amssymb}
\usepackage{xcolor}
\definecolor{lcolor}{rgb}{0.5,0,0}
\definecolor{citcolor}{rgb}{0,0.3,0.0}
\usepackage[breaklinks,colorlinks,urlcolor=blue,citecolor=citcolor,linkcolor=lcolor]{hyperref}

\newlength{\fullfigwidth}
\newcommand{\lqcd}{\Lambda_{\mathrm{QCD}}}
\newcommand{\nc}{N_{\mathrm{c}}}
\newcommand{\qs}{Q_{\mathrm{s}}}
\newcommand{\as}{\alpha_{\mathrm{s}}}
\newcommand{\ud}{\, \textrm{d}}
\newcommand{\tr}{\textrm{Tr}}
\newcommand{\nr}[1]{(\ref{#1})}
\newcommand{\eq}{Eq.~}

\begin{document}
\title{Spectral function for overoccupied gluodynamics from classical lattice simulations
\thanks{Presented at the XXV Epiphany conference on Advances in Heavy Ion Physics}%
}
\author{
K. Boguslavski
\address{
Institut f{\"u}r Theoretische Physik, Technische Universit{\"a}t Wien, 1040 Vienna, Austria
\\ and \\
Department of Physics 
 P.O. Box 35, 40014 University of Jyv\"askyl\"a, Finland
}
\\ \rule{0pt}{0pt}
\\
A. Kurkela
\address{
Theoretical Physics Department, CERN,
CH-1211 Geneva, Switzerland 
\\ and
\\  
Faculty of Science and Technology, University of Stavanger, 4036 Stavanger, Norway
}
\\  \rule{0pt}{0pt}
\\
T. Lappi
\address{
Department of Physics, %
 P.O. Box 35, 40014 University of Jyv\"askyl\"a, Finland
\\ and \\
Helsinki Institute of Physics, P.O. Box 64, 00014 University of Helsinki,
Finland
}
\\  \rule{0pt}{0pt}
\\
J.  Peuron
\address{
European Centre for Theoretical Studies in Nuclear Physics and Related Areas (ECT*) and
Fondazione Bruno Kessler, Strada delle Tabarelle 286, I-38123 Villazzano (TN), Italy
}
}

\maketitle

\begin{abstract}
We study the spectral properties of an overoccupied gluonic system far from equilibrium. Using classical Yang-Mills  simulations and linear response theory, we determine the statistical and spectral functions. We measure  dispersion relations and damping rates of transversally and longitudinally polarized excitations in the gluonic plasma, and also study further structures in the spectral function. 
\end{abstract}
  
\section{Introduction}

The main purpose of the program of ultrarelativistic heavy ion collisions is to create and study the properties of deconfined QCD matter in the laboratory.  Our purpose here is to study strongly overoccupied color field configurations that are relevant for several of the different aspects of the collision process.  In the very initial pre-equilibrium stage after the collision the  dynamics is dominated by  \emph{gluon saturation}. The characteristic aspect of this regime is the existence of a semihard dominant transverse momentum scale, the saturation scale $\qs \gg \lqcd$, generated by nonlinear interactions of the dense gluonic system. At the saturation scale the gluon field is nonperturbatively strong. This means that the gluon field strengths, or occupation numbers of gluonic states, are parametrically proportional to the inverse of the coupling $A_\mu A_\mu\sim 1/\as$. Later in the evolution it is generally believed that the plasma reaches a state close to local thermal equilibrium. In such a thermal system a part of the degrees of freedom, namely the soft fields with momenta $p \lesssim gT$ are, similarly, nonperturbatively large. These soft fields are important for many  properties of QCD matter.  Thus in both cases we are in a situation where we want to understand the \emph{real time} behavior of QCD systems with both a  perturbative momentum scale (generically denoted here by $Q\gg\lqcd$) and therefore weak coupling constant $\as \ll 1$, but also gluon fields (at least for some important momentum modes) in an  overoccupied state. In this regime the approximation of \emph{classical fields} provides a powerful nonperturbative tool.

The standard method for understanding real time QCD dynamics in weak coupling is provided by the hard (thermal) loop (HTL) approach. Here one develops a perturbation theory based on a separation of two different momentum scales. The degrees of freedom at the hard scale  $Q$ can be thought of as (classical) particles, and interact with soft ($\sim m_D$) modes that can be thought of as (classical) fields. In an equilibrium plasma  the small coupling constant provides such a scale separation, but this approach can also be generalized to nonequilibrium systems. In addition to analytical calculations, there are also many numerical implementations (see e.g.~\cite{Hu:1996sf,Moore:1997sn,Bodeker:1999gx})  of this idea, based on  explicitly different descriptions of particle- and field-like degrees of freedom. Such calculations have been used to understand e.g. sphaleron transitions in thermal systems and plasma instabilities in anisotropic ones. 

In a heavy ion collision the system starts from a configuration, at  very early times $\tau \sim  1/\qs$, where there is only one scale $Q\sim m_D \sim \qs$. Equilibration is, then, the process where the two scales develop through various stages to become parametrically different, $m_D \ll T$. Following such a scenario, especially at its earliest stages, is problematic with a qualitatively different description of hard and soft modes. In practical terms, if the soft modes are described as fields on a lattice with lattice spacing $a$, the lattice needs to be fine enough to represent them: $m_D \ll 1/a$. But a  particle-like physical picture of the  hard  modes implies that they are localized at the scale of the lattice spacing, requiring $a \gg 1/Q$ since the de Broglie wavelength of the hard particles is $\sim 1/Q$. In such a description it is therefore not possible to have $m_D\sim Q$. 

The approach that we are following here avoids this problem. We treat all degrees of freedom on the same lattice, subject to the same lattice UV cutoff $1/a$. Thus we do not \emph{need} to have a large scale separation.  On the other hand, classical lattice simulations being relatively inexpensive, one can  fit in a rather large separation between the hard and soft scales, even while maintaining a controllable continuum limit $Q\ll 1/a$. We treat even the hard modes as classical fields, not particles. In the HTL limit the classical treatment of the hard scales should not matter for the hard+soft interactions, since the only important thing about the hard modes are their color currents, not whether these currents are made of particles or fields. A drawback of the classical lattice approach is that the interactions between hard modes are treated incorrectly (as classical fields instead of particles as they should), although in the overoccupied regime the error is of higher order in $\as$.  Thus our system would, ultimately, thermalize to an unphysical classical equilibrium. The interactions between the hard modes are, however, much slower, and often neglected in the particle+field simulations in any case. We refer the reader to e.g. Refs.~\cite{Kurkela:2012hp,Berges:2013lsa,York:2014wja} for a discussion on the validity of the classical approximation. One can see our calculational setup as an extension of HTL setup to situations where the scale separation $m_D/ Q$ can be varied smoothly up to large values.

This talk reviews the recent results presented in more detail in Ref.~\cite{Boguslavski:2018beu}.
In the following we shall briefly describe our numerical method of linearized fluctuations, based on the algorithm developed in Ref.~\cite{Kurkela:2016mhu} and the obervables that we measure. We then introduce our test case system, the isotropic self-similar UV cascade of gluons. We will then review some of our numerical results so far and discuss interesting prospects for future projects. 

\section{Methods}

For the time evolution of Classical Yang-Mills fields we use the standard Hamiltonian lattice~\cite{Wilson:1974sk,Kogut:1974ag} formulation of gauge theory in real time. Here, instead of the gauge potential $A_i$ and the  covariant derivative $D_i=\partial_i+ig[A_i,\cdot]$, one works with link matrices connecting lattice sites
$U_i(x)= e^{iagA_i(x)}$ and covariant finite  differences obtained using the links. The canonical conjugate variable to the gauge potential is the chromoelectric field $E^i = \partial_t A_i$. The Hamiltonian setup is formulated in the temporal gauge $A_0=0$, where one must take care to satisfy also the Gauss' law constraint  $[D_i,E^i]=0$.

Our first measurable is  the ``statistical function,''  defined as a symmetric two point function of the classical field
\begin{equation}
 \label{eq_stat_fct}
 F_{jk}^{ab} (x, x') = \frac{1}{2}\left\langle \left\{ \hat{A}_j^a(x) \, , \, \hat{A}_k^b(x') \right\} \right\rangle .
\end{equation}
The statistical function measures ``thermal'' fluctuations in the field, and is related to the phase space density of  particles in system $f(p)$. In the classical approximation the commutator is suppressed by a power of $g$ (because $A_i\sim 1/g$ ). Thus here $F$ is just a two point function of the classical field
\begin{equation}
F_{jk}^{ab} (x,x') = \left\langle A_j^b(x) A_k^b(x') \right\rangle_{\mathrm{cl}}.
\end{equation}
The statistical function at equal time is an often measured quantity in CYM calculations, but in Ref.~\cite{Boguslavski:2018beu} we also measure it at unequal time.

In addition to the statistical function, the other in general independent correlator in quantum field theory is the  ``spectral function'' 
\begin{equation}
\rho_{jk}^{ab} (x, x') = i \left\langle \left[ \hat{A}_j^a(x), \hat{A}_k^b(x') \right] \right\rangle .
\end{equation}
It is a genuinely ``quantum'' obervable and proportional to  $\sim \hbar$ due to the the field operators commutation relations. We can nevertheless measure it in the CYM simulations by using its relation to the retarded propagator as
\begin{equation} 
G_{R,jk}^{ab}(t,t',p) = \theta(t - t')\, \rho_{jk}^{ab}(t,t',p). 
\end{equation}
The retarded propagator measures the linear response of the system to an external perturbation, and we measure it with the algorithm developed in Ref.~\cite{Kurkela:2016mhu}.  We split the gauge field into a background field and a fluctuation:
\begin{equation}
\hat{A}_i^a(x) \to \hat{A}_i^a(x) +  \hat{a}_i^a(x) ,
\end{equation}
where the fluctuation generated by an external infinitesimal current is given bt the retarded propagator
\begin{equation}
\langle \hat{a}_i^b(x)\rangle = \int \ud^4 x' G_{R,ik}^{~~bc}(x, x')\, j^k_c(x').
\end{equation}
This leads to our algorithm to calculate the statistical function.
We first, at time $t_0$, perturb  the system with a  current that exists only for one timestep: $j^k_c(x) \sim  e^{i \mathbf{k}\cdot \mathbf{x}} \delta(t-t_0)$. We then follow the  linearized equations of motion for the fluctuation and its time derivative $a_i^a(x) = \langle \hat{a}_i^a(x) \rangle,\ e^i_a(x)$. Finally, at a later time $t>t_0$, we
 measure the correlation of the  field $a_i^a(t)$ with the  current $j^i_a(t_0)$ and use this to extract the momentum space spectral function   $\rho(p,t)$. A similar procedure with the electric field fluctuation can be used to obtain time derivatives of the spectral function.

\section{Overoccupied cascade}

Let us then move on to discuss our test case overoccupied nonequilibrium system, the isotropic self-similar  cascade. This system has been extensively studied by many groups (see e.g. \cite{Berges:2012ev,Kurkela:2012hp,York:2014wja}), and it has been found that a kinetic theory formulation can describe the basic properties seen in the numerical studies. In this system one starts from an isotropic field configuration where only modes up to some characteristic hard momentum are occupied
\begin{equation}
f(p) \sim \frac{n_0}{g^2}\theta(p_0-p).
\end{equation}
In practical calculations one typically replaces a strict theta function with a smoother Gaussian momentum distribution, but this detail, or the initial occupation number $n_0/g^2 \sim 1/g^2$ or momentum scale $p_0$  matter little for the later self-similar time evolution of the system. This is instead fully determined by the conserved total energy density, which defines the real characteristic hard scale of the problem as   $\varepsilon\sim Q^4/g^2$. In what follows we specifically define  $Q = \sqrt[4]{ 10 \pi^{3/2} g^2 \varepsilon/\left(\sqrt{2}(\nc^2-1)\right) }$, 
scale dimensionful quantities with $Q$ to dimensionless ones and, unless otherwise stated, plot quantities measured at $Qt=1500$.

\begin{figure}[htb]
\centerline{%
\includegraphics[width=6cm]{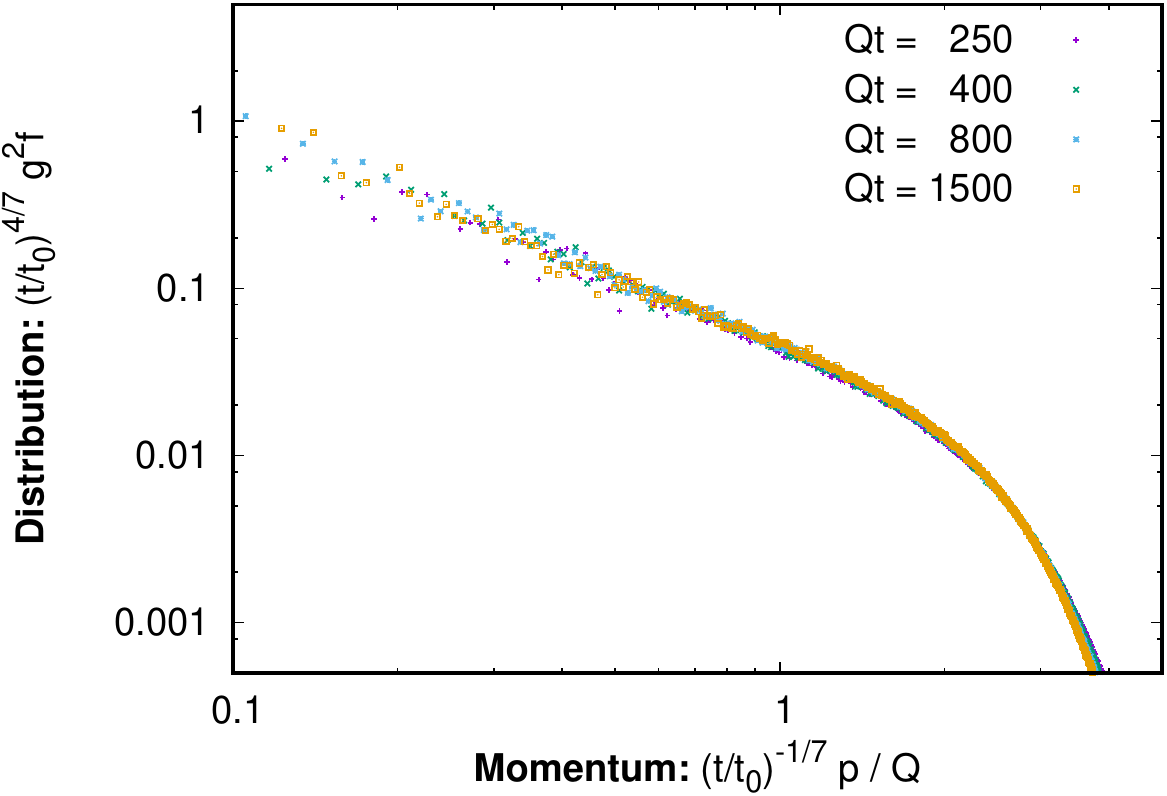}
\hfill
\includegraphics[width=6cm]{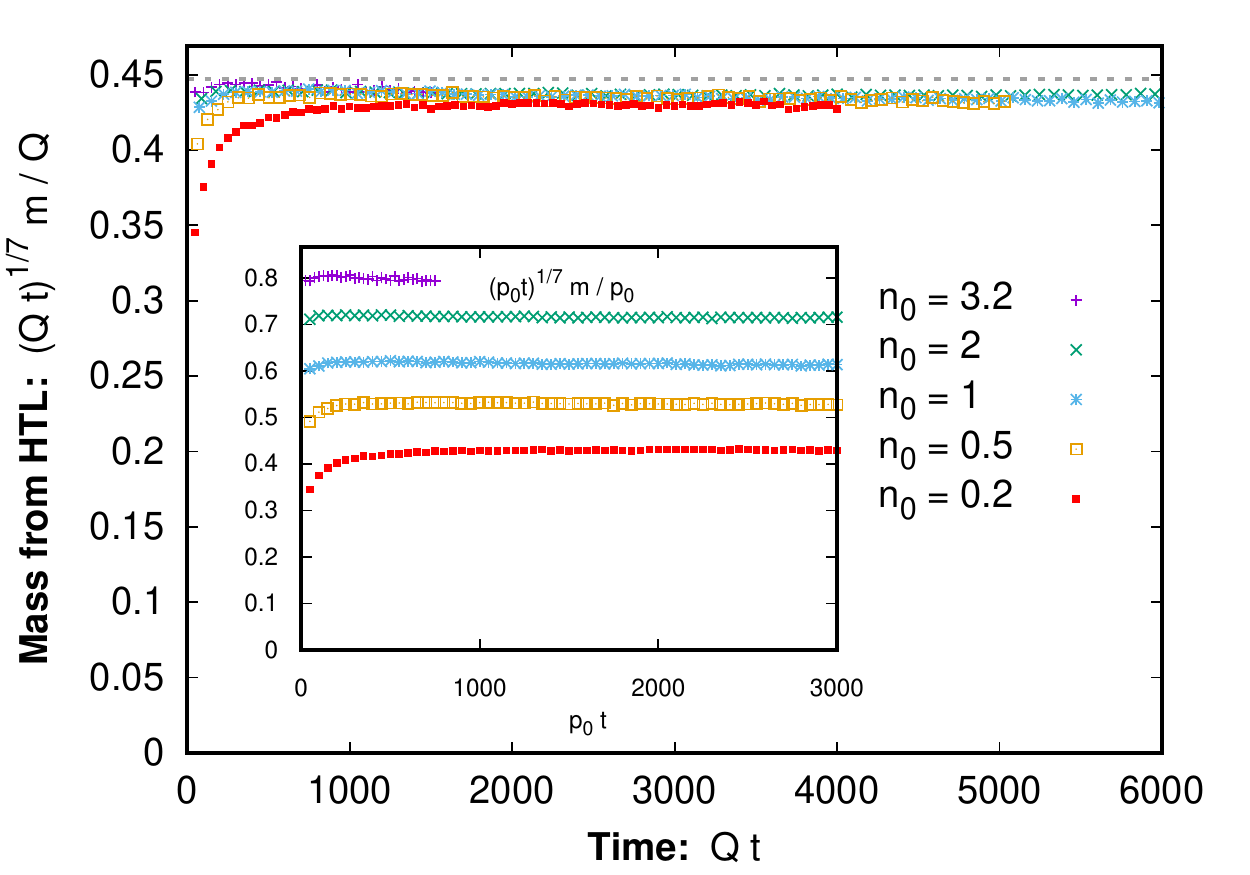}
}
\caption{Left: Scaled momentum distribution in terms of the rescaled momentum, demonstrating the self-similar nature of the cascade. Right: Scaling behavior of the Debye scale, with the inset showing the unscaled values. The distribution $f(p)$ is estimated using the equal time electric field correlator.}
\label{fig:scaling}
\end{figure}

During the time evolution of the system, the energy cascades towards UV modes in such a way that all the momentum modes up to  $p_\mathrm{max}\sim t^{1/7}$ are occupied. The typical occupation number (e.g. at the hard scale $p_\mathrm{max}$) decreases with time as\begin{equation}
f(p_\mathrm{max}) \sim t^{-4/7}.
\end{equation}
The scaling behavior 
\begin{equation} \label{eq:scaling}
 f(t,p) = t^{-4/7} f_\mathrm{s}(p/t^{1/7} )
\end{equation}
becomes evident in the numerical result when we scale the momentum distribution of gluons by $t^{4/7}$, and plot it in terms of the scaled momentum $t^{-1/7}p$, as shown in Fig.~\ref{fig:scaling} (left).

In HTL or kinetic theory one estimates the Debye or plasmon scale from the (hard) particle distribution as
\begin{equation}
\label{eq:htlmd}
m^2 = 2 \nc \int \frac{\ud^3 \mathbf{p}}{(2\pi)^ 3} \frac{f(p)}{p} .
\end{equation}
In the scaling regime~\nr{eq:scaling} this gives an estimate 
\begin{equation}
m \sim t^{-1/7}
\end{equation}
for the time dependence of the soft scale. This scaling is verified numerically in Fig.\ref{fig:scaling} (right).  This leads us to an important aspect of this self-similar attractor system, namely that the scale separation $p_\mathrm{max}/m$ grows with time. By looking at different times one can therefore smoothly turn on or off the parameter that determines the validity of the HTL approximation. In the following we will be mostly working at rather late times where this scale separation is clear, in order to compare our calculation to HTL in a regime where it should be a good description of the system.

\section{Spectral function and dispersion relations}

\begin{figure}[htb]
\centerline{%
\includegraphics[width=4cm]{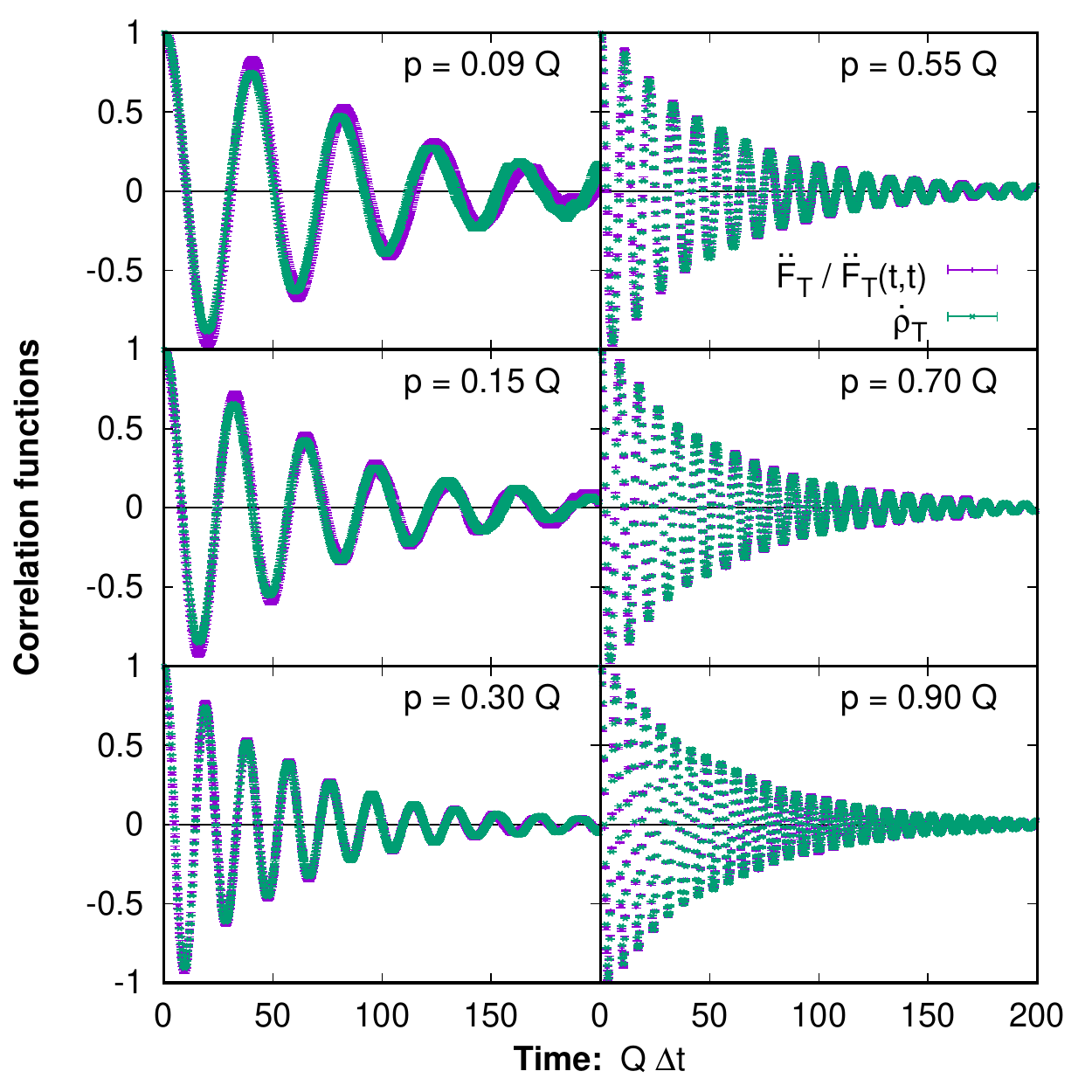}
\hfill
\includegraphics[width=4cm]{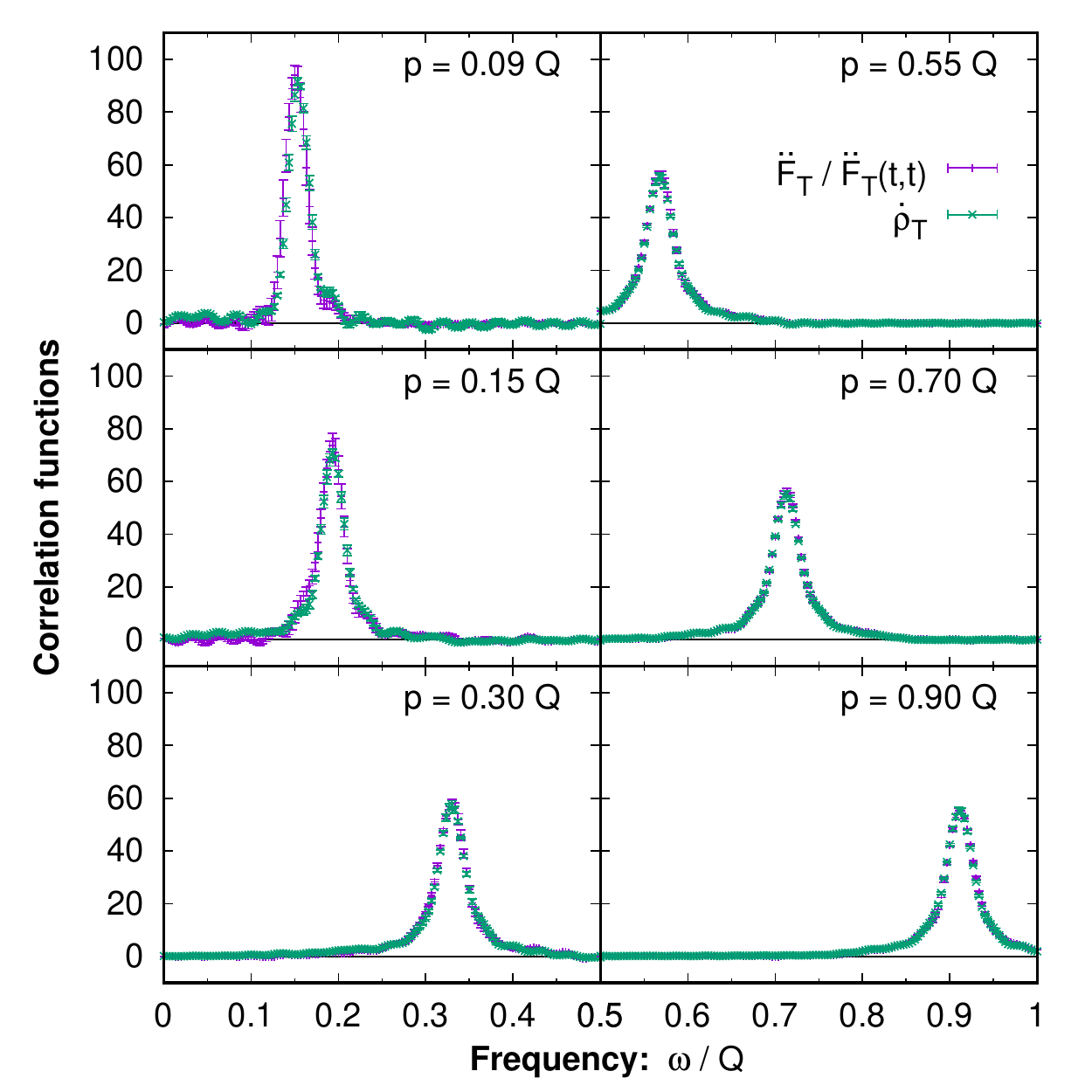}
\hfill
\includegraphics[width=4cm]{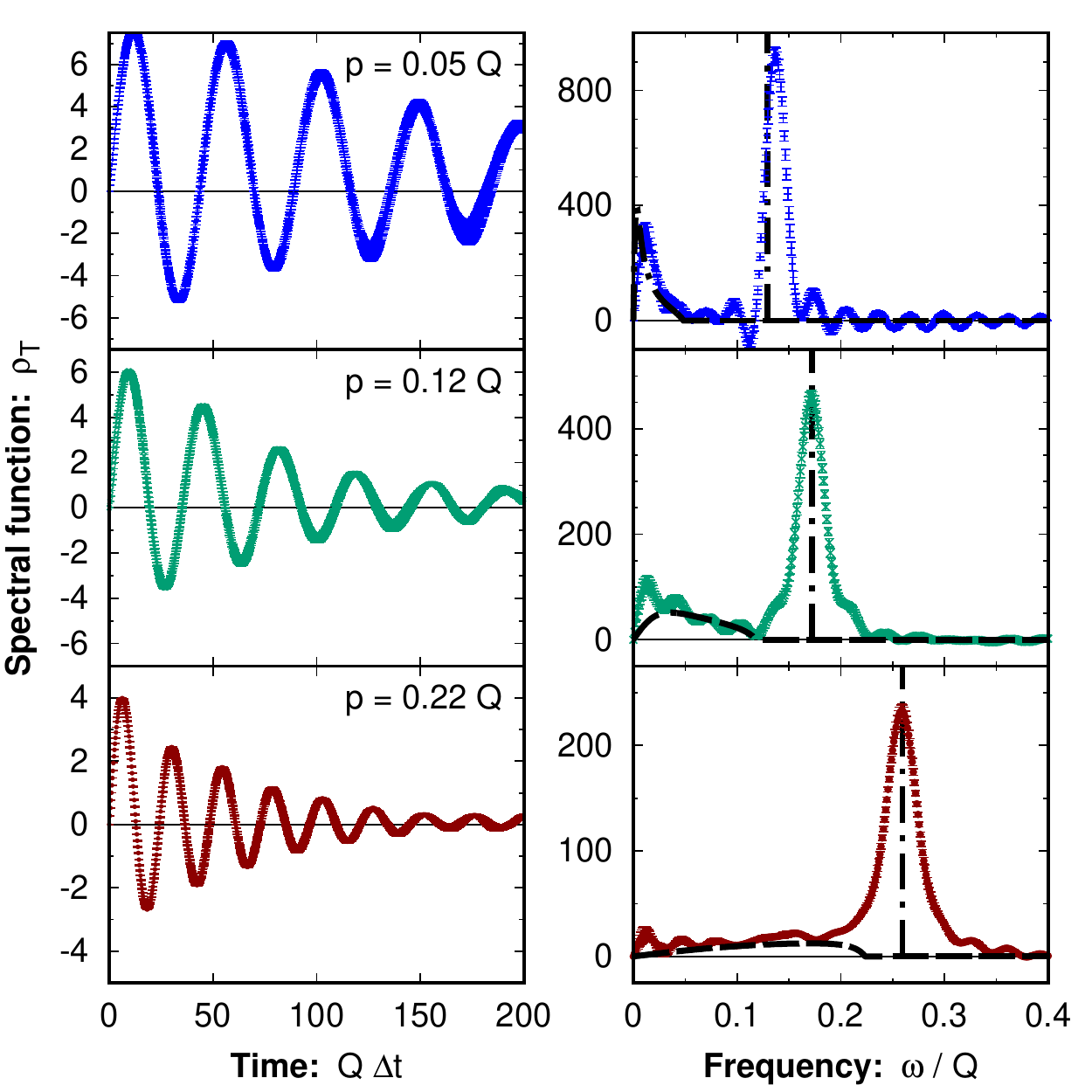}
}
\caption{Left: spectral and statistical functions in the time domain for transverse polarization. Center: the same functions in the frequency domain, corresponding to $\omega \rho(\omega)$. Here the statistical function has been divided by its equal time value, which makes the normalization the same at zero time separation $\Delta t=0$. Right: the spectral function $\rho$ in the time and frequency domain. With one power of frequency less, the structures at small $\omega$ are now more visible. The dashed black line is the LO HTL functional form, exhibiting a zero-width peak and a ``Landau cut'' region at small $\omega$.} 
\label{fig:rhot}
\end{figure}

We start this discussion with the statistical and spectral functions for transversely polarized quasiparticles. Relevant questions here are whether they both exhibit the same peak structure, and what are the locations and widths of the peaks. At small frequencies one would also expect to see an additional structure, the ``Landau cut''.  The spectral function is naturally normalized to unity (actually $\hbar$), whereas the statistical function is proportional to the number of particles in the system. In order to compare the peak structure in the two correlators it is convenient to divide the statistical function by its value at $\Delta t = t'-t=0$, which is what we will do in the following plots. Figure \ref{fig:rhot} shows the spectral function for the transverse polarization. First of all we see that there is a very nice agreement between the two functions, which could be interpreted as a generalized fluctuation-dissipation theorem. In frequency space one can see a very nice Lorentzian shape, and extract the quasiparticle width (plasmon damping rate). One can even clearly see the Landau cut structure at small frequencies, in rough agreement with the HTL theory curve that uses only the value of $m$ extracted numerically from the data as an input parameter.

\begin{figure}[htb]
\centerline{%
\includegraphics[width=4cm]{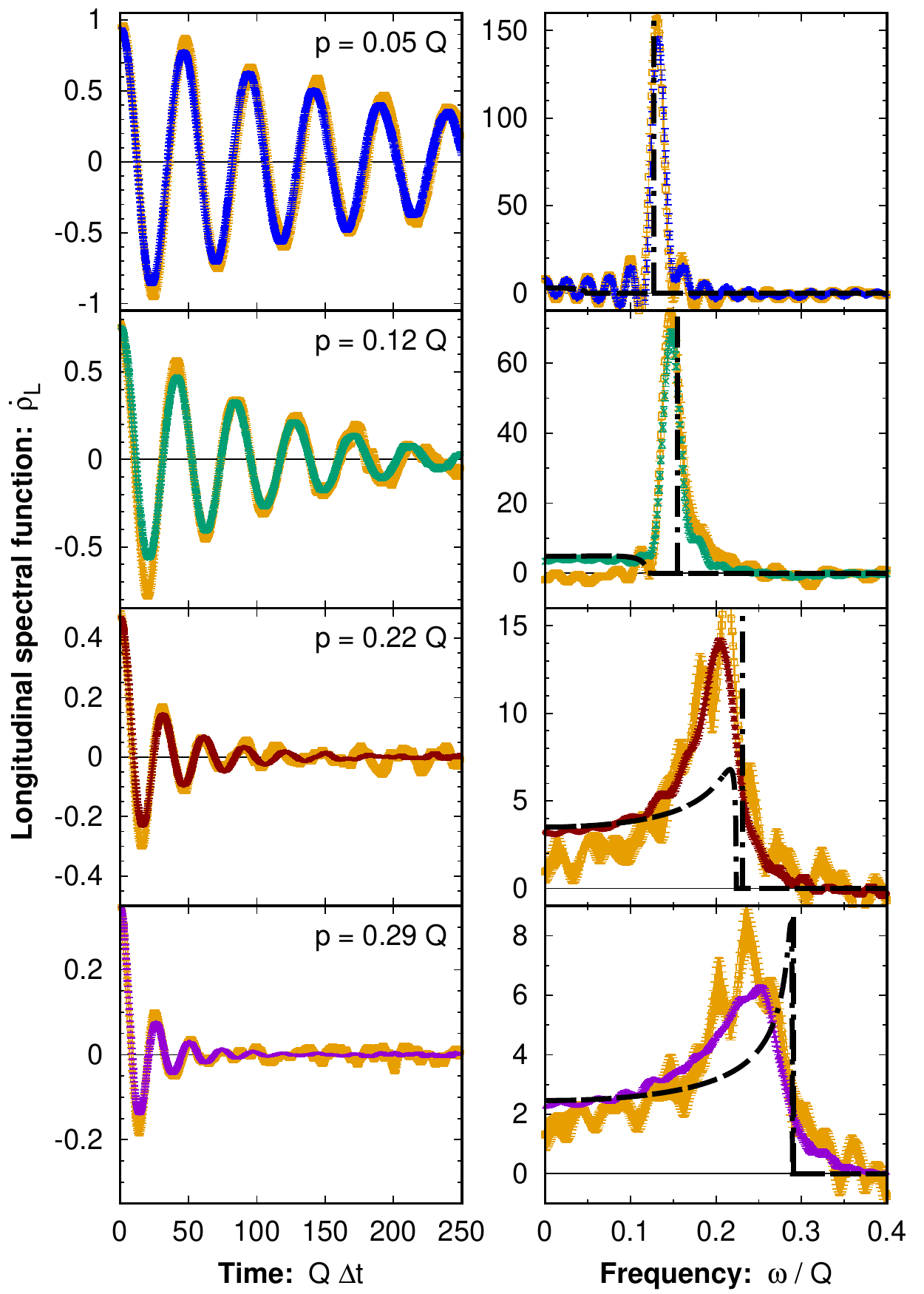}
}
\caption{Statistical (orange) and spectral (other colors) function for the longitudinal polarization in the time and frequency domains. 
} 
\label{fig:rhol}
\end{figure}

For the longitudinal polarization mode, as shown in Fig.~\ref{fig:rhol} the  story is very similar.  There is a  good agreement between the statistical and spectral functions, when the former is normalized to the equal time value. For the longitudinal polarization state the measurement is harder, since at high momenta the quasiparticle peak gets weaker and merges with the Landau cut, as could have been expected from HTL.

\begin{figure}[htb]
\centerline{%
\includegraphics[width=6cm]{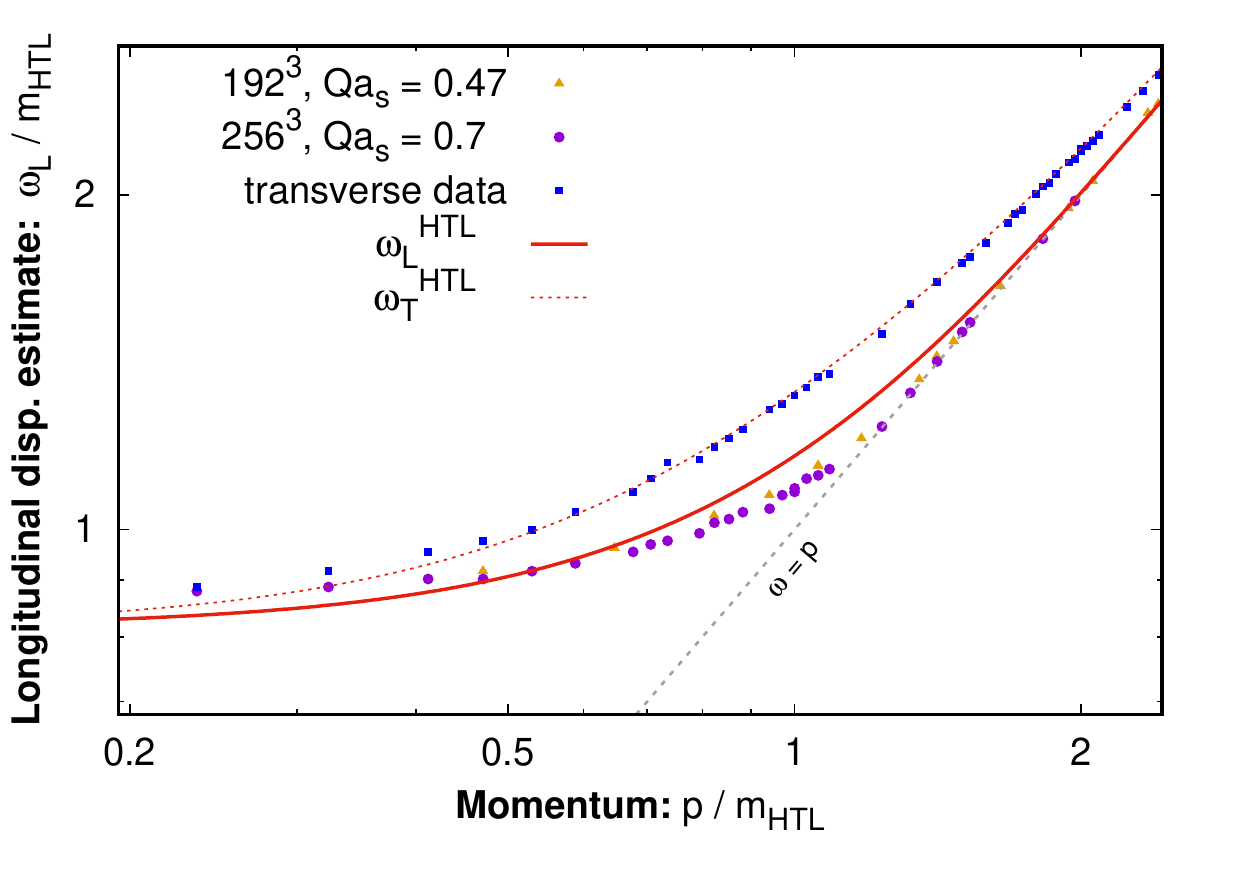}
\hfill 
\includegraphics[width=6cm]{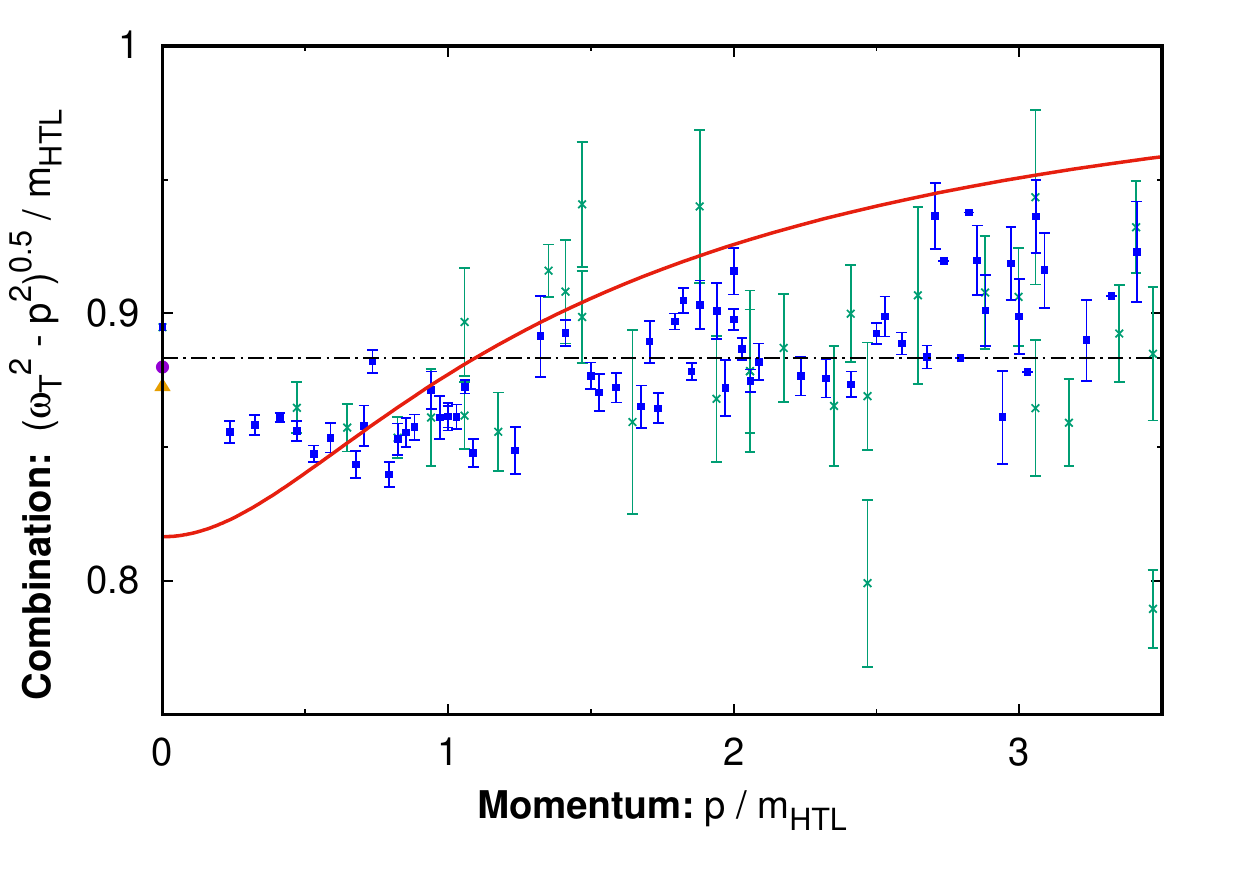}
}
\caption{Left: dispersion relation for transverse and longitudinal polarizations, compared to the HTL functional form using $m$ calculated from $f(p)$ using \eq\nr{eq:htlmd}
and to a relativistic dispersion relation $\omega^2 = m^2+p^2$. Right: $\sqrt{\omega^2-p^2}/m$ vs $p$ to better illustrate the difference between a relativistic dispersion relation (dashed straight line), the HTL functional form (solid red line) and the numerical result for the transverse polarization state. The greeen crosses are a simulation with  $Qa=0.47$ and the blue stars with $Qa=0.7$.} 
\label{fig:disprel}
\end{figure}

By extracting the location of the peak as a function of momentum, we can extract a quasiparticle dispersion relation from our data as shown in Fig.~\ref{fig:disprel}. Overall, within our statistics it is not possible to differentiate between the HTL functional form and a relativistic dispersion relation $\omega^2 = m^2+p^2$. Figure~\ref{fig:disprel} also shows data for the longitudinal polarization. The difference between the two polarization states is qualitatively as one would expect from HTL.
Quantitatively we can characterize the dispersion relation by two different masses from the small and large momentum limits by defining the plasmon mass $\omega_\mathrm{pl} \equiv \omega(p \to 0)$ and the mass gap at  $p\to \infty$, denoted by  $ m_\infty$. Our numerical estimate for the ratio of these scales is
\begin{equation}
\frac{ \omega_\mathrm{pl} }{m_\infty} = 0.96 ,
\end{equation}
where HTL prediction for this quantity would be
\begin{equation}
\frac{ \omega_\mathrm{pl} }{m_\infty} = \sqrt{2/3} \approx 0.82,
 \end{equation}
and the NLO correction has been calculated to be negative~\cite{Schulz:1993gf}.

\section{Further observables}
\begin{figure}[htb]
\centerline{%
\includegraphics[width=6cm]{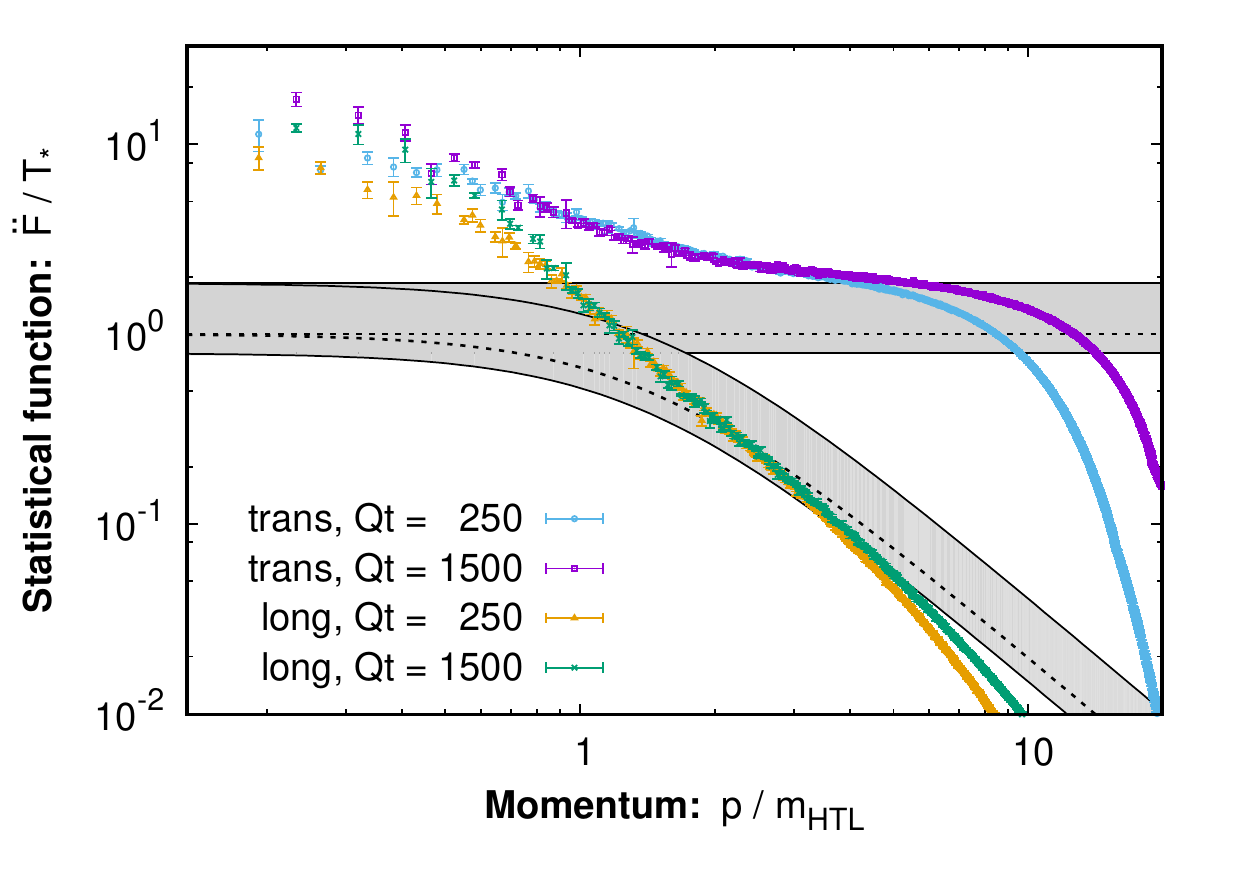}
\hfill
\includegraphics[width=6cm]{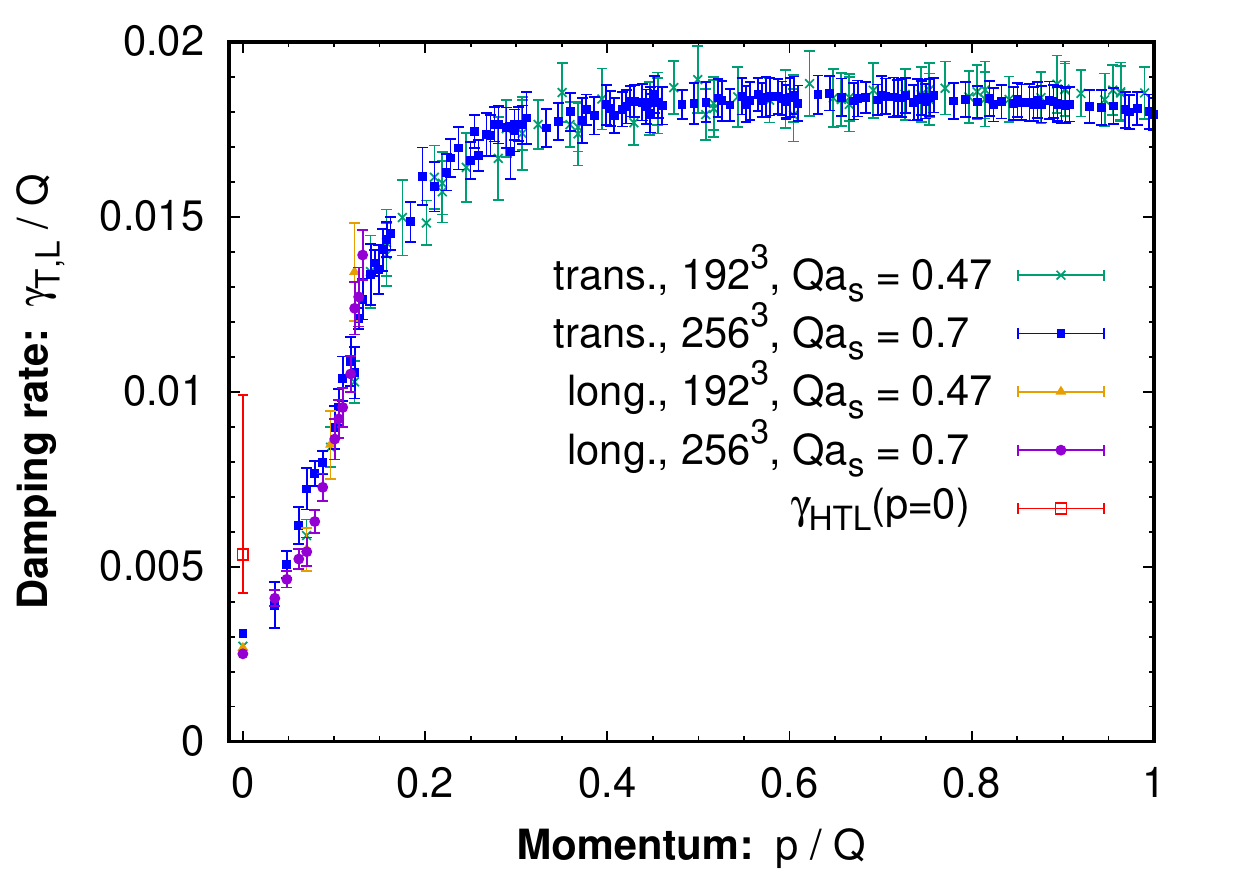}
}
\caption{Left: The equal time electric field correlator for transverse and longitudinal polarizations. The shaded gray bands are the expectation from HTL theory. Right:
The plasmon damping rate. From HTL theory we only have an estimate for the value at $p=0$.} 
\label{fig:ee}
\end{figure}

Let us now discuss a little more the equal time field correlator, i.e. $f(p)$, which so far was used as a normalization factor to compare the statistical and spectral functions. In the small momentum limit the expectation from HTL theory would be to see a classical thermal distribution 
\begin{equation}
f(p)\approx \frac{T_*}{\omega(p)} .
\end{equation}
The effective temperature  for the soft fields should be given by 
\begin{equation}\label{eq:tstar}
T_* \equiv \frac{\frac{1}{2} \int_{\mathbf{p}} f(t,p) \left(f(t,p) + 1 \right)
 }{
 \int_{\mathbf{p}} \frac{f(t,p)}{\sqrt{m^ 2 + p^2}}}
 \sim t^{-3/7},
 \end{equation}
where for  classical fields one  neglects the  1 in $(f+1)$. Our results for the equal time electric field correlator are shown in Fig.~\ref{fig:ee} (left). We see that there is a significant enhancement in the infrared compared to the expectation. We do not currently have a compelling interpretation for this observation, but certainly the agreement with expectations from HTL is not as good as for the spectral function.

In the HTL theory the plasmon damping rate is also related to the effective temperature $T_*$. Our numerical result for the damping rate, extracted from fitting the time domain signal with an exponentially decaying function,  is shown in Fig.~\ref{fig:ee} (right). Our measurements of the time dependence of $\gamma$ are consistent with the time scaling from \eq\nr{eq:tstar}. However, we are able to extract a result in a rather large range of momenta, whereas from the HTL theory we only have points at $p=0$  (shown on the figure) and $p=\infty$~\cite{Pisarski:1993rf} (not shown, but consistent within errors). Also here, the agreement of this $T_*$-dependent observable with HTL is not as good as for the spectral function.

\section{Future prospects and conclusions}

After this quick review of the first results in Ref.~\cite{Boguslavski:2018beu}, let us briefly discuss what we see as natural next steps in this setup. Given this powerful formalism for real-time calculations the natural next step would be to use it for calculations of transport coefficients in such a nonequilibrium gluon plasma. One would be interested in e.g. the heavy quark diffusion coefficient, the shear viscosity $\eta$ or the jet quenching parameter $\hat{q}$. Probably the simplest of these is the first one. Without going into details (see e.g. \cite{Moore:2004tg}), for our purposes it is enough to state that this is defined as the infinite-time limit of the local unequal time electric field correlator:
\begin{equation}
\kappa \equiv \int_0^\infty\ud t \ \kappa(t),   \quad \textrm{where } \kappa(t) \propto \tr \left< E^i(0,\mathbf{x}) E^i(t,\mathbf{x})\right>.
 \end{equation}
Here we have omitted the temporal gauge links needed to make this quantity explicitly gauge-invariant, since they become unity in temporal gauge. The correlator is in principle a straighforward quantity, but surprisingly tricky to evaluate numerically. This is due to the fact that the integrand is very strongly oscillatory, and very fine cancellations are needed to obtain the value of the integral. However, the time dependence of the integrand itself contains more physical information than just the constant. In fact, our preliminary studies strongly suggest that the time dependence contains very strong slowly oscillating modes. These could be a confirmation of the infrared enhancement seen in the gauge fixed correlator in Fig.~\ref{fig:ee}. The correlator $\kappa(t)$ is manifestly gauge invariant, unlike the electric field correlators as a function of $p$ that are evaluated in Coulomb gauge.  Thus the time dependence could serve as a gauge invariant confirmation of the IR enhancement. 

\begin{figure}[t]
\centerline{%
\includegraphics[width=\fullfigwidth]{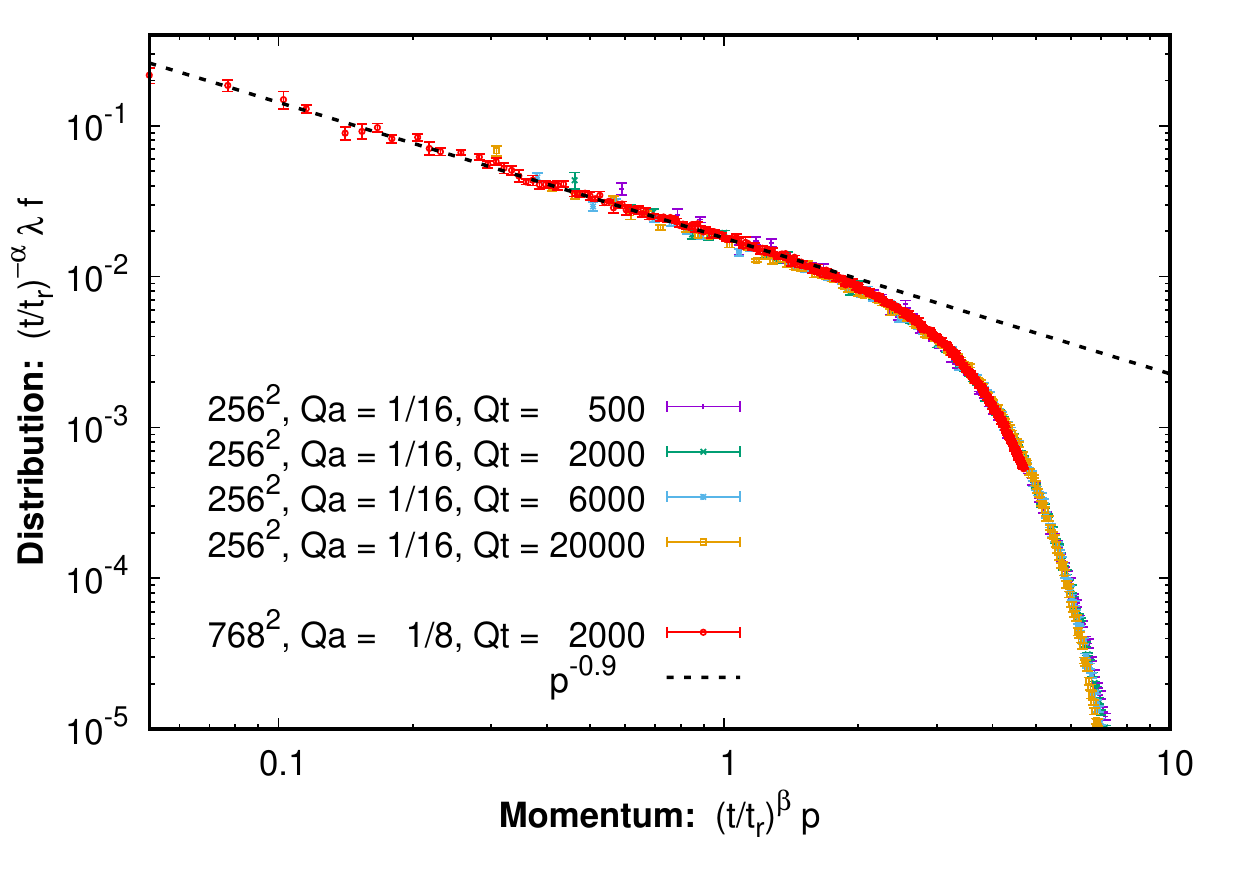}
}
\caption{The 2-dimensional scaled particle number distribution as a function of the scaled momentum, at different times and for different lattice spacings, at
$Q t = 500$, with the scaling line parametrized by $\alpha = 3\beta,$ $\beta = -1/5.7$. 
}
\label{fig:2dscaling}
\end{figure}

Another natural step is to move from a 3-dimensional isotropic cascade towards the initial stage of a relativistic heavy ion collision at high energy. Here the system starts off, at leading order in the QCD coupling, as an effectively 2-dimensional  boost invariant glasma~\cite{Lappi:2006fp} field. In a genuinely 2-dimensional system it seems that an HTL-type approximation cannot be applied. One way to see this is to look at the integral \nr{eq:htlmd} that gives the value of the Debye mass in terms of the distribution of hard particles, e.g. in thermal equilibrium. In 3 spatial dimensions the integral is dominated by the hard modes, and thus the Debye mass is in a sense an ``external'' parameter for the soft fields. In 2 spatial dimensions, assuming that the distribution in the IR is close to thermal $f\sim 1/\omega$, the integral \nr{eq:htlmd} receives a large, logarithmic, contribution from soft modes. Thus the Debye mass is not an external parameter anymore, but must be self-consistently determined from the soft modes. Thus it is not obvious whether a kinetic theory description exists for the 2-dimensional system, and specifically whether one would expect to see a self-similar cascade solution as a function of time. This is a situation that can perfectly well be studied in our setup of classical fields and linearized fluctuations. Concentrating at this stage on a nonexpanding 2-dimensional  system (which was studied already in \cite{Lappi:2017ckt}, but only at early times and with equal-time correlators), we find clear evidence for a self-similar cascade solution, whether or not this has a kinetic theory interpretation. Our preliminary numerical result for the distribution function is shown in Fig.~\ref{fig:2dscaling}.

In conclusion, we have here argued that several aspects of a heavy ion collision exhibit overoccupied $f(p) \sim 1/g^2$ classical gauge field configurations. This includes the initial glasma fields  with the only relevant momentum scale $p\sim \qs$, and, later, the soft fields $p\sim gT$ in thermal system.  For controlled understanding of these fields we have developed a  new numerical algorithm for linearized fluctuations on top of the fully nonlinearly interacting classical gauge field. As a first  test case of this setup we study an isotropic self-similar UV cascade solution. At late enough times, there exists a scale separation between the hard and soft modes. Thus one can compare the results to HTL. We have seen that the HTL approximation is rather well satisfed for the spectral function, but more uncertain for quantities related to the effective temperature of the soft modes, denoted as $T_*$. In particular, we are able to extract a plasmon decay rate $\gamma(p)$ as a function of momentum. In the future, we see many prospects in extending these methods to transport coefficients, with the heavy quark diffusion coefficient as a first step. We also want to study anisotropic systems relevant to the pre-equilibrium matter in heavy ion collisions, where as a  first application we demostrate a self-similar scaling solution in a purely 2-dimensional gauge theory.

\subsection*{Acknowledgements}
This work is supported  by the European Research Council, grant ERC-2015-CoG-681707. T.~L.\ has been supported by the Academy of Finland, projects No. 267321 and No. 303756.  J.~P.\ has been supported by the Jenny and Antti Wihuri Foundation and a travel grant from the Magnus Ehrnrooth foundation. K.~B.\ and J.~P.\ would like to thank the CERN Theory group for hospitality during part of this work. We acknowledge computational resources from CSC – IT Center for Science, Finland.

\bibliographystyle{h-physrev4mod2}
\bibliography{spires}

\providecommand{\href}[2]{#2}\begin{thebibliography}{10}

\bibitem{Hu:1996sf}
C.~R. Hu and B.~Muller,
\newblock \href{http://dx.doi.org/10.1016/S0370-2693(97)00851-4}{Phys. Lett.
  {\bf B409}, 377 (1997)},
  [\href{http://arXiv.org/abs/hep-ph/9611292}{{arXiv:hep-ph/9611292
  [hep-ph]}}].

\bibitem{Moore:1997sn}
G.~D. Moore, C.-r. Hu and B.~Muller,
\newblock \href{http://dx.doi.org/10.1103/PhysRevD.58.045001}{Phys. Rev. {\bf
  D58}, 045001 (1998)},
  [\href{http://arXiv.org/abs/hep-ph/9710436}{{arXiv:hep-ph/9710436
  [hep-ph]}}].

\bibitem{Bodeker:1999gx}
D.~Bodeker, G.~D. Moore and K.~Rummukainen,
\newblock Phys. Rev. {\bf D61}, 056003 (2000),
  [\href{http://arXiv.org/abs/hep-ph/9907545}{{arXiv:hep-ph/9907545}}].

\bibitem{Kurkela:2012hp}
A.~Kurkela and G.~D. Moore,
\newblock \href{http://dx.doi.org/10.1103/PhysRevD.86.056008}{Phys. Rev. {\bf
  D86}, 056008 (2012)}, [\href{http://arXiv.org/abs/1207.1663}{{arXiv:1207.1663
  [hep-ph]}}].

\bibitem{Berges:2013lsa}
J.~Berges, K.~Boguslavski, S.~Schlichting and R.~Venugopalan,
\newblock \href{http://dx.doi.org/10.1007/JHEP05(2014)054}{JHEP {\bf 05}, 054
  (2014)}, [\href{http://arXiv.org/abs/1312.5216}{{arXiv:1312.5216 [hep-ph]}}].

\bibitem{York:2014wja}
M.~C. Abraao~York, A.~Kurkela, E.~Lu and G.~D. Moore,
\newblock \href{http://dx.doi.org/10.1103/PhysRevD.89.074036}{Phys. Rev. {\bf
  D89}, 074036 (2014)}, [\href{http://arXiv.org/abs/1401.3751}{{arXiv:1401.3751
  [hep-ph]}}].

\bibitem{Boguslavski:2018beu}
K.~Boguslavski, A.~Kurkela, T.~Lappi and J.~Peuron,
\newblock \href{http://dx.doi.org/10.1103/PhysRevD.98.014006}{Phys. Rev. {\bf
  D98}, 014006 (2018)},
  [\href{http://arXiv.org/abs/1804.01966}{{arXiv:1804.01966 [hep-ph]}}].

\bibitem{Kurkela:2016mhu}
A.~Kurkela, T.~Lappi and J.~Peuron,
\newblock \href{http://dx.doi.org/10.1140/epjc/s10052-016-4523-9}{Eur. Phys. J.
  {\bf C76}, 688 (2016)},
  [\href{http://arXiv.org/abs/1610.01355}{{arXiv:1610.01355 [hep-lat]}}].

\bibitem{Wilson:1974sk}
K.~G. Wilson,
\newblock \href{http://dx.doi.org/10.1103/PhysRevD.10.2445}{Phys. Rev. {\bf
  D10}, 2445 (1974)}.

\bibitem{Kogut:1974ag}
J.~B. Kogut and L.~Susskind,
\newblock \href{http://dx.doi.org/10.1103/PhysRevD.11.395}{Phys. Rev. {\bf
  D11}, 395 (1975)}.

\bibitem{Berges:2012ev}
J.~Berges, S.~Schlichting and D.~Sexty,
\newblock \href{http://dx.doi.org/10.1103/PhysRevD.86.074006}{Phys. Rev. {\bf
  D86}, 074006 (2012)}, [\href{http://arXiv.org/abs/1203.4646}{{arXiv:1203.4646
  [hep-ph]}}].

\bibitem{Schulz:1993gf}
H.~Schulz,
\newblock \href{http://dx.doi.org/10.1016/0550-3213(94)90624-6}{Nucl. Phys.
  {\bf B413}, 353 (1994)},
  [\href{http://arXiv.org/abs/hep-ph/9306298}{{arXiv:hep-ph/9306298
  [hep-ph]}}].

\bibitem{Pisarski:1993rf}
R.~D. Pisarski,
\newblock \href{http://dx.doi.org/10.1103/PhysRevD.47.5589}{Phys. Rev. {\bf
  D47}, 5589 (1993)}.

\bibitem{Moore:2004tg}
G.~D. Moore and D.~Teaney,
\newblock \href{http://dx.doi.org/10.1103/PhysRevC.71.064904}{Phys. Rev. {\bf
  C71}, 064904 (2005)},
  [\href{http://arXiv.org/abs/hep-ph/0412346}{{arXiv:hep-ph/0412346
  [hep-ph]}}].

\bibitem{Lappi:2006fp}
T.~Lappi and L.~McLerran,
\newblock \href{http://dx.doi.org/10.1016/j.nuclphysa.2006.04.001}{Nucl. Phys.
  {\bf A772}, 200 (2006)},
  [\href{http://arXiv.org/abs/hep-ph/0602189}{{arXiv:hep-ph/0602189}}].

\bibitem{Lappi:2017ckt}
T.~Lappi and J.~Peuron,
\newblock \href{http://dx.doi.org/10.1103/PhysRevD.97.034017}{Phys. Rev. {\bf
  D97}, 034017 (2018)},
  [\href{http://arXiv.org/abs/1712.02194}{{arXiv:1712.02194 [hep-lat]}}].

\end{thebibliography}

\end{document}